\documentclass[10pt,3p,authoryear]{elsarticle}
\usepackage{tgtermes}
\usepackage[dvipsnames]{xcolor}%colors
\usepackage{amsmath,amsfonts,amsthm,amssymb}
\usepackage{lineno,hyperref}
\hypersetup{
  colorlinks=true,
  pdftitle={Morphological stability of electrostrictive thin films},
}
\usepackage{geometry}
\usepackage{fancyhdr}
\usepackage{siunitx}
\usepackage[figurename=Fig.,labelfont=bf,labelsep=period]{caption}
\providecommand{\dod}[3][]{\dfrac{\mathrm{d}^{#1}#2}{\mathrm{d}{{#3}^{#1}}}}
\providecommand{\dpd}[3][]{\dfrac{\partial^{#1}#2}{\partial^{{#1}}{{#3}^{#1}}}}
\providecommand{\dfd}[3][]{\dfrac{\delta{^{#1}}#2}{\delta^{{#1}}{{#3}^{#1}}}}
\providecommand{\intj}[4][]{\int_{#2}^{#1} \! {#3} \, \mathrm{d} {#4}}
\newcommand{\iu}{\mathrm{i}\mkern1mu} % imaginary symbol
\newcommand{\abs}[2][0]{\vert {#2} \vert}
\graphicspath{{./figures/}}% path to figures and images
\begin{document}

% header and footer
\pagestyle{fancy} % Set the page style to "fancy"
\fancyhf{} % sets both header and footer to nothing
\renewcommand{\headrulewidth}{0pt} % no header rule
\fancyhead{} % clear all header fields
\fancyhead[R]{\footnotesize{\textit{International Journal of Engineering Science}}}
\fancyhead[L]{\footnotesize{\textit{J. Zhang and P.W. Voorhees}}}
\fancyfoot{} % clear all footer fields
\fancyfoot[C]{\small{\thepage}}

\begin{frontmatter}

\title{Morphological stability of electrostrictive thin films}

\author[nwu]{Jin Zhang}\corref{cor1}
\author[nwu]{Peter W. Voorhees}
\cortext[cor1]{jzhang@northwestern.edu}
\affiliation[nwu]{
organization={Department of Materials Science and Engineering},
addressline={Northwestern University},
city={Evanston},
postcode={60208},
state={IL},
country={USA}
}

\begin{abstract}
  A large electric field is typically present in anodic or passive oxide films. Stresses induced by such a large electric field are critical in understanding the breakdown mechanism of thin oxide films and improving their corrosion resistance. In this work, we consider electromechanical coupling through the electrostrictive effect. A continuum model incorporating lattice misfit and electric field-induced stresses is developed. We perform a linear stability analysis of the full coupled model and show that, for typical oxides, neglecting electrostriction underestimates the film's instability, especially in systems with a large electric field. Moreover, a region where electrostriction can potentially provide a stabilizing effect is identified, allowing electrostriction to enhance corrosion resistance. We identified an equilibrium electric field intrinsic to the system and the corresponding equilibrium film thickness. The film's stability is very sensitive to the electric field: a 40 percent deviation from the equilibrium electric field can change the maximum growth rate by nearly an order of magnitude. Moreover, our model reduces to classical morphological instability models in the limit of misfit-only, electrostatic-only, and no-electrostriction cases. Finally, the effect of various parameters on the film's stability is studied.
\end{abstract}

\begin{keyword}
Electromechanical processes \sep Oxidation \sep Corrosion \sep Breakdown \sep Thin film
\end{keyword}

\end{frontmatter}

%% ===========================================================================

\section{Introduction}\label{sec:intro}

The formation of passive oxide film provides corrosion resistance to most industrially important metals and alloys. Passive films generally are very thin (a few nanometers thick). The electric fields in the film have been observed to be on the order of $10^8-10^9~\si{\volt\per\meter}$. Stresses induced by such a large field are expected to be important in the breakdown of the passive film (\cite{Sato1971,Heuer2011,Heuer2012}). The piezoelectric effect is one of the electromechanical coupling effects. However, it only exists in non-centrosymmetric materials. In contrast, electrostriction is an electromechanical coupling effect that applies to all crystal symmetries. Experimental observations by \cite{Heuer2011} and \cite{Heuer2012} show that electrostrictive stress can be on the same order of magnitude as deformation-induced stresses and is critical in the growth of passive films.

Theoretical modeling of thin-film oxidation with electromechanical coupling is vital in understanding the breakdown mechanism of thin oxide films. Stresses associated with the breakdown of oxide films consist of the stress due to lattice misfit between the substrate and the film and the stress induced by the electrical field, including the Maxwell stress (\cite{Suo2008}) and the electrostrictive stress (\cite{Heuer2011}). 

Many models that describe the growth of oxide films are based on the thermodynamics and kinetics of point defects, \textit{e.g.} \cite{Atkinson1985,Macdonald1990,Seyeux2013,Ramanathan2019}. However, the effects of stress are often neglected. The first work that accounts for the electric field-induced stress is the work of \cite{Sato1971}. The Maxwell stress (mistermed as electrostrictive stress) on the growth of passive films is considered but not the electrostrictive stress. However, experimentally shown by \cite{Heuer2012}, the electrostrictive stress can be much larger than the Maxwell stress in a thin film; thus, it is not negligible. \cite{Tang2011} perform morphological stability analysis with electrostriction. However, several assumptions are made so the system is not fully coupled: the perturbation of the electric field within the film is decoupled from the mechanical and electrical equations, and the misfit strain is accounted for in a simplified way where the interaction between the film and the substrate is neglected. \cite{Stanton2008} consider the instability with electrostriction but neglect the misfit strain and do not utilize the full electromechanical coupling.

This work presents a continuum model with full electromechanical coupling to describe the growth of oxide films, considering the effect of lattice misfit, Maxwell stress, and electrostriction. The coupled mechanical and electrical equations are solved analytically, and linear morphological stability analysis is performed. The equilibrium electric field, an intrinsic system property, is identified. We show that our model reduces to previous models under certain limits. It is shown that the electric field has a large effect on the stability, and the electrostrictive effect can dominate, especially with a large electric field. Finally, we studied the impact of various parameters, including the interfacial energy, the electric field, the oxide's permittivity, the misfit strain, and the shear modulus of the oxide and the substrate.

\textit{Notation.} We use Einstein notation with summation from 1 to 3 over repeated indices, and the order of tensor can be determined from the number of indices. In a few cases, we use bold letters for vectors. $(\cdot)_{,i}$ denotes the partial derivative to the $i$-th spatial coordinate, $(\cdot)_{,t}$ denotes the time derivative, and $\delta_{ij}$ denotes the Kronecker delta. The bar $\bar{\cdot}$ and hat $\hat{\cdot}$ symbols indicate the basic state (planar interface) and the perturbed quantities, respectively.

\section{Theory}\label{sec:method}
In this section, we formulate our theory, which couples deformation and electric response by electromechanical coupling through electrostriction. The fundamental equations are described in Section~\ref{sec:method:model}. The basic-state (planar interface) solutions are provided in Section~\ref{sec:method:basic}, and the linear stability analysis (perturbed interface) is given in Section~\ref{sec:method:stability}.

\subsection{Problem formulation}\label{sec:method:model}

\begin{figure}[!t]
\centering
\includegraphics[width=.72\textwidth]{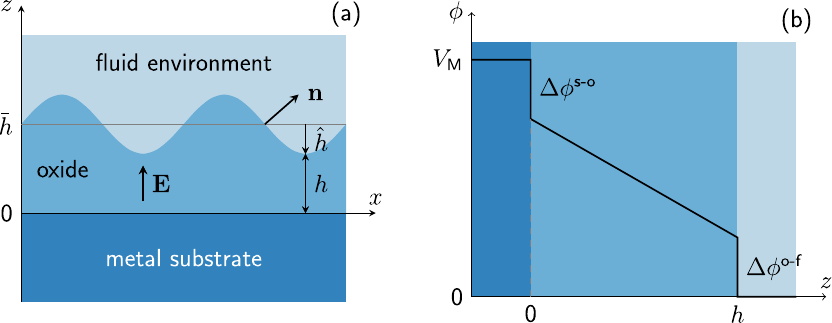}
\caption{The substrate-oxide-fluid system: (a) a schematic diagram of the system; (b) profile of the electric potential along a cross-section of constant $x$.}\label{fig:oxidation}
\end{figure}
The problem considered in this paper is shown in Fig.~\ref{fig:oxidation}a. An oxide film forms on a metal substrate and in contact with a liquid or vapor environment. An electric field $\mathbf{E}$ exists inside the oxide film because of charge separation across the oxide film. We use the Cabrera-Mott model (\cite{Cabrera1949}), which assumes the oxide film is thin enough that electrons from the metal can tunnel through the film to participate in a reduction reaction on the surface of the oxide. The oxide film is then treated charge neutral. The metal substrate and the fluid environment are field-free, as shown in Fig.~\ref{fig:oxidation}b. The metal substrate has a constant electrostatic potential $\phi=V_{\text{M}}$, which can be due to externally applied voltage or self-corrosion. The lattice mismatch between the substrate and the oxide, which can be substantial, leads to a misfit strain in the film (\cite{Yu2018}). No shear stress exists in the fluid phase. The substrate-oxide interface is assumed to be coherent for simplicity. The growth of oxide film is assumed to be limited by the chemical reaction at the oxide-fluid interface, allowing us to neglect detailed point defect structures and focus on the effect of electromechanical coupling. Mechanical and electrical relaxation are assumed to be much faster than the interface evolution; therefore, the system is in mechanical and electrical equilibrium. The oxide-fluid interface evolves to reduce the system's total free energy, including chemical free energy, elastic strain energy, and electrostatic energy. The substrate-oxide interface is planar and assumed to be stationary. The coordinate system is depicted in Fig.~\ref{fig:oxidation}. $z=0$ represents the substrate-oxide interface, and $z=h(x,y,t)$ represents the oxide-fluid interface. We assume all the field variables are homogeneous along the $y$-coordinate; therefore, the $y$-dependence is discarded in the rest of the paper.

\subsubsection{Field variables}
Based on the description of the system given above, the system's status can be uniquely determined by the following state variables: the displacements $u_i^\alpha$ and the electrostatic potential $\phi^\alpha$ of each phase and the location of the oxide surface $h(x,t)$. Here, $\alpha=\text{s,o,f}$ denotes the substrate, oxide, and fluid phases. Other field variables can then be determined from these independent state variables.

The total linear elastic strain tensor is related to the displacement $u_i$ as
\begin{equation}\label{eq:strain}
\varepsilon_{ij} = \frac{1}{2}(u_{i,j}+u_{j,i}).
\end{equation}
The electric field is $E_i = - \phi_{,i}$. The Cauchy stress in the passive film is
\begin{equation}\label{eq:stress}
\sigma_{ij} = C_{ijkl} (\varepsilon_{kl}-\varepsilon^{\text{p}}_{kl}),
\end{equation}
where $C_{ijkl}$ is the stiffness tensor and $\varepsilon^{\text{p}}_{kl}$ is the eigenstrain (\cite{Voorhees2004}). The reference state for strain measurement is defined as the zero-field stress-free state of the substrate. Note that this is the reference state for both the substrate and the oxide. In the presence of lattice misfit and electric field, the eigenstrain for the oxide is $\varepsilon^{\text{p}}_{kl}=\varepsilon^{\text{misfit}}_{kl}+\varepsilon^{\text{es}}_{kl}$. The misfit strain of a cubic crystal is $\varepsilon^{\text{misfit}}_{kl} = \varepsilon^* \delta_{kl}$ with $\varepsilon^*=(a^{\text{o}}-a^{\text{s}})/a^{\text{s}}$, where $a^{\text{o}}$ and $a^{\text{s}}$ are the lattice constants of the oxide and the substrate, respectively. The electrostrictive strain is a function of the electric field: $\varepsilon^{\text{es}}_{kl}=M_{klmn}E_mE_n$, where $M_{klmn}$ is a fourth-order tensor of the electrostrictive constants (\cite{Heuer2011}). The electrostrictive stress can then be written as $\sigma_{ij}^{\text{es}}=C_{ijkl}M_{klmn}E_mE_n=\frac{1}{2}l_{ijmn}E_mE_n$. Here, $l_{ijmn}$ is the electrostrictive tensor.

\subsubsection{Mechanical and electrical equilibrium in bulk phases}

Mechanical equilibrium in each phase is given by (\cite{Dorfmann2005})
\begin{equation}\label{eq:equilibrium:mechanical}
(\sigma_{ij}^{\alpha} + \sigma^{\text{M},\alpha}_{ij})_{,j} = 0,\quad\alpha=\text{s,o,f}.
\end{equation}
The Maxwell stress is
\begin{equation}\label{eq:sigma:Maxwell}
\sigma^{\text{M}}_{ij} = \epsilon E_iE_j - \frac{1}{2} \epsilon E_kE_k \delta_{ij},
\end{equation}
where $\epsilon$ is the permittivity of the corresponding phase. Since the substrate and fluid phases are field-free, the Maxwell stress is only present in the oxide.

Electrical equilibrium in the oxide phase is given by
\begin{equation}\label{eq:equilibrium:electrical}
D_{i,i}^{\text{o}} = 0,
\end{equation}
where $D_i$ is the electric displacement. Since $E_i^{\text{s}} = 0$ and $E_i^{\text{f}}=0$, we only need to consider the electrical equilibrium in the oxide. Note here that we assume the film, substrate, and fluid are all charge neutral, and there is no body force.

\subsubsection{Boundary conditions}

Boundary conditions are needed to uniquely solve the governing equations given above. We require a continuity of the normal traction on the metal-oxide and oxide-fluid interfaces:
\begin{equation}\label{eq:bc:mechanical:so}
\sigma_{i3}^{\text{o}}+\sigma_{i3}^{\text{M,o}} = \sigma_{i3}^{\text{s}}\quad \text{at }z=0,
\end{equation}
\begin{equation}\label{eq:bc:mechanical:of}
(\sigma_{ij}^{\text{o}}+\sigma_{ij}^{\text{M,o}}) n_j = 0\quad \text{at }z=h(x,t),
\end{equation}
where $n_j$ is the unit normal vector of the oxide-fluid interface pointing from the oxide to the fluid (see Fig.~\ref{fig:oxidation}a). Since the pressure in the fluid phases is typically orders of magnitude smaller than the stress in the oxide, the stress in the fluid is assumed to be zero: $\sigma^{\text{f}}_{ij}=0$. At the metal-oxide interface, the displacements are continuous
\begin{equation}\label{eq:bc:mechanical:u:so}
u_i^{\text{o}}(z=0) = u_i^{\text{s}}(z=0).
\end{equation}
The elastic strain in the metal far from the interface decays to zero
\begin{equation}\label{eq:bc:mechanical:u:s}
\varepsilon_{ij}^{\text{s}} \rightarrow 0 \quad \text{as }z\rightarrow-\infty.
\end{equation}
The electric potential on the fluid side of the oxide-fluid interface is chosen as the zero reference $\phi^\text{f}(h)=0$. The electric potential in the metal substrate $V_{\text{M}}$ is a constant. The potential jumps across each interface are denoted as $\Delta \phi^{\text{s-o}}$ and $\Delta \phi^{\text{o-f}}$, see Fig.~\ref{fig:oxidation}b. For simplicity, both potential jumps are assumed to be constant in this work. The boundary conditions on the electric potential in the oxide are
\begin{equation}\label{eq:bc:electrical:phi}
\phi^{\text{o}}(z=0)=V_{\text{M}}-\Delta \phi^{\text{s-o}},\quad \phi^{\text{o}}(z=h)=\Delta \phi^{\text{o-f}}.
\end{equation}

\subsubsection{Interfacial boundary condition}

The velocity of the oxide-fluid interface (defined to be positive for growing oxide) due to a reaction at the interface follows a linear reaction model
\begin{equation}\label{eq:interfacial}
v_n=-L \mathcal{F},
\end{equation}
where $L$ is a kinetic parameter, $\mathcal{F}$ is the driving force expressed as (see \ref{sec:appendix:interfacial} for the derivation):
\begin{equation}\label{eq:F}
\mathcal{F} = \Delta g_{v} + n_i D_i E_j n_j + \gamma \kappa,%
\end{equation}
where $\gamma$ is the interfacial energy of the oxide-fluid interface, $\kappa=n_{i,i}$ is the total curvature, and $\Delta g_v=g_v^{\text{o}}-g_v^{\text{f}}$ is the difference in the electric Gibbs free energy density between the oxide and the fluid. The electric Gibbs free energy density of phase $\alpha$ is written as
\begin{equation}\label{eq:gv}
g_v^{\alpha} = f_{\text{chem}}^{\alpha} + W_v^{\alpha},
\end{equation}
where $f_{\text{chem}}^{\alpha}$ is the chemical free energy density, and $W_v^{\alpha}$ is a free energy density due to stress and electric field. We have $\Delta g_v=\Delta f_{\text{chem}}+\Delta W_v$, where $\Delta f_{\text{chem}}=f_{\text{chem}}^{\text{o}}-f_{\text{chem}}^{\text{f}}$. Note that $\mathcal{F}$ is evaluated at the oxide-fluid interface, \textit{i.e.}, $z=h(x,t)$. The normal velocity and the total curvature are given by the interface profile $h(x,t)$ as $v_n = h_{,t}/(1+h_{,x}^2)^{\frac{1}{2}}$ and $\kappa=-h_{,xx}/(1+h_{,x}^2)^{\frac{3}{2}}$, respectively.

\subsubsection{Constitutive equations}

Finally, constitutive relations are needed to complete the equations. We first look at the constitutive equations for the oxide. The chemical free energy $f_{\text{chem}}^{\text{o}}$ is assumed independent of strain and electric field. The free energy density $W_v^{\text{o}}$ is (\cite{Tang2011})
\begin{equation}\label{eq:Wv}
W_v^{\text{o}} = \frac{1}{2} C_{ijkl}(\varepsilon_{ij}-\varepsilon^* \delta_{ij}) (\varepsilon_{kl}-\varepsilon^*\delta_{kl}) - \frac{1}{2} \epsilon E_k E_k - \frac{1}{2} l_{ijkl} E_i E_j (\varepsilon_{kl}-\varepsilon^* \delta_{kl}).
\end{equation}
Here, we use linear isotropic material models:
\begin{equation}\label{eq:mat:linearisotropic}
\begin{split}
C_{ijkl} &= \mu (\delta_{ik}\delta_{jl}+\delta_{il}\delta_{jk}) + \lambda \delta_{ij}\delta_{kl},\\
l_{ijkl} &= \frac{1}{2} a_1 (\delta_{ik}\delta_{jl}+\delta_{il}\delta_{jk}) + a_2 \delta_{ij}\delta_{kl},
\end{split}
\end{equation}
where $\mu$ and $\lambda$ are Lam\'e constants, and $a_1$ and $a_2$ are constants related to electrostriction. In this work, we use the model by \cite{Shkel1998}
\begin{equation}\label{eq:a1a2}
\begin{array}{l}
\displaystyle a_1 = -\frac{2}{5} (\epsilon_r-1)^2 \epsilon_0,\\
\displaystyle a_2 = \left(-\frac{1}{3}(\epsilon_r-1)(\epsilon_r+2)+\frac{2}{15}(\epsilon_r-1)^2\right)\epsilon_0,
\end{array}
\end{equation}
where $\epsilon_r$ is the relative permittivity, and $\epsilon_0$ is the vacuum permittivity. The permittivity is $\epsilon = \epsilon_r \epsilon_0$. In the limit of no electrostriction, we have $a_1=a_2=0$. Note that if $\epsilon_r=1$, there is no electrostrictive effect from Eq.~\ref{eq:a1a2}.

The Cauchy stress in the oxide film is derived from $g_v^{\text{o}}$ as
\begin{equation}\label{eq:sigma}
\sigma_{ij}^{\text{o}} = \dpd{g_v^{\text{o}}}{\varepsilon_{ij}} = 2\mu \varepsilon_{ij} + \lambda \varepsilon_{kk}\delta_{ij} -(2\mu+3\lambda)\varepsilon^*\delta_{ij} - \frac{1}{2}\left(a_1 E_i E_j + a_2 E_kE_k \delta_{ij}\right).
\end{equation}
The last term accounts for the electrostrictive effect. Eq.~\ref{eq:sigma} is consistent with Eq.~\ref{eq:stress}. Note that Maxwell stress in Eq.~\ref{eq:sigma:Maxwell} can be written as part of the derivative of the electric Gibbs free energy in a large deformation formulation; see \cite{Suo2008}. In other words, the derivative gives $\sigma_{ij}^{\text{o}}+\sigma^{\text{M}}_{ij}$ in a large deformation formulation. Here, we adopt the small strain formulation.

The electric displacement in the oxide film is derived as
\begin{equation}\label{eq:D}
D_i^{\text{o}} = -\dpd{g_v^{\text{o}}}{E_i} = \left(\epsilon \delta_{ij} + a_1 \varepsilon_{ij} + a_2 \varepsilon_{kk}\delta_{ij} - (a_1+3a_2)\varepsilon^*\delta_{ij}\right)E_j.
\end{equation}
It can be seen that electrostriction acts as a modification of the equivalent permittivity.

For the substrate, the Maxwell stress and the electric displacement are both zero: $\sigma_{ij}^{\text{M,s}}=0$ and $D_i^{\text{s}}=0$. The Cauchy stress in the substrate is
\begin{equation}\label{eq:sigma:sf}
\sigma^{\text{s}}_{ij} = 2\mu^{\text{s}} \varepsilon^{\text{s}}_{ij} + \lambda^{\text{s}} \varepsilon^{\text{s}}_{kk}\delta_{ij}.
\end{equation}

For the fluid, all field quantities are zero.

In summary, the displacements $u_i^{\text{o}}$ and $u_i^{\text{s}}$ and the electric potential $\phi^{\text{o}}$ are solved from Eqs.~\ref{eq:equilibrium:mechanical} and \ref{eq:equilibrium:electrical}, together with the constitutive relations in Eqs.~\ref{eq:sigma}-\ref{eq:sigma:sf} and the boundary conditions given by Eqs.~\ref{eq:bc:mechanical:so}-\ref{eq:bc:electrical:phi}. The evolution of the oxide-fluid interface is then determined from the interfacial boundary condition in Eq.~\ref{eq:interfacial}. For simplicity, we neglect the superscript $(\cdot)^{\text{o}}$ of quantities for oxide in the rest of the paper.

\subsection{Basic state}\label{sec:method:basic}

Now, we proceed to solve the systems of equations given in Section~\ref{sec:method:model}. We first consider a planar interface and derive its solutions, regarded as the basic-state solution. Consider the thickness of the planar film as $\bar{h}$. At the basic state, the metal substrate is fully relaxed: $\bar{u}_i^{\text{s}} = 0$. In the oxide film, the basic-state displacements and electric potential are solved as $\bar{u}_1=\bar{u}_2=0$, $\bar{u}_3 = \bar{\varepsilon}_{33} z$, and
$\bar{\phi} =\Delta \phi^{\text{o-f}} + \bar{E}_3 (\bar{h} - z)$, where
\begin{equation}\label{eq:basic:epsilon33}
\bar{\varepsilon}_{33}= \frac{2\mu+3\lambda}{2\mu+\lambda}\varepsilon^* - \frac{(\epsilon - a_1 - a_2)}{2 (2\mu+\lambda)}\bar{E}_3^2,
\end{equation}
\begin{equation}\label{eq:basic:E3}
\bar{E}_3 = \frac{V_{\text{M}}-\Delta\phi^{\text{s-o}}-\Delta\phi^{\text{o-f}}}{\bar{h}}.
\end{equation}
The basic-state strain in Eq.~\ref{eq:basic:epsilon33} includes two contributions: one related to the misfit strain $\varepsilon^*$ and one induced by the electric field.

The stresses in the oxide are then determined from the constitutive relations:  $\bar{\sigma}_{12}=\bar{\sigma}_{13}=\bar{\sigma}_{23}=0$, 
\begin{equation}\label{eq:basic:stress11}
\bar{\sigma}_{11}=\bar{\sigma}_{22} = - \frac{2\mu(2\mu+3\lambda)}{2\mu+\lambda}\varepsilon^* - \frac{\lambda}{2\mu+\lambda} \frac{\epsilon}{2} \bar{E}_3^2 - \left(\frac{a_2}{2}-\frac{\lambda(a_1+a_2)}{2(2\mu+\lambda)}\right) \bar{E}_3^2,
\end{equation}
$\bar{\sigma}_{33} = -\epsilon \bar{E}_3^2/2$, $\bar{\sigma}_{12}^{\text{M}}=\bar{\sigma}_{13}^{\text{M}}=\bar{\sigma}_{23}^{\text{M}}=0$, $\bar{\sigma}_{11}^{\text{M}}=\bar{\sigma}_{22}^{\text{M}}=-\epsilon \bar{E}_3^2/2$, and $\bar{\sigma}_{33}^{\text{M}} = \epsilon \bar{E}_3^2/2$. The basic state stress in Eq.~\ref{eq:basic:stress11} includes three contributions: the first is related to the misfit strain, the second is related to the Maxwell stress, and the last is related to electrostriction. \\
The electrical displacement in the oxide is $\bar{D}_1=\bar{D}_2=0$,
\begin{equation}\label{eq:basic:D3}
\bar{D}_3=\left(\epsilon - (a_1+3a_2)\varepsilon^*+(a_1+a_2)\bar{\varepsilon}_{33}\right)\bar{E}_3.
\end{equation}
Both strain $\bar{\varepsilon}_{33}$ and misfit strain $\varepsilon^*$ contribute to the electric displacement. The coupling between mechanical deformation and electric response is bidirectional and becomes clear from Eqs.~\ref{eq:basic:epsilon33} and \ref{eq:basic:D3}: strain can be induced by the electric field, and polarization is affected by strain. Note that in the absence of electrostriction, $a_1=0$ and $a_2=0$, the strain does not contribute to the electric displacement, while the electric field still contributes to the strain through Maxwell stress, see Eq.~\ref{eq:basic:epsilon33}. 

\begin{figure}[!t]
\centering
\includegraphics[width=.8\textwidth]{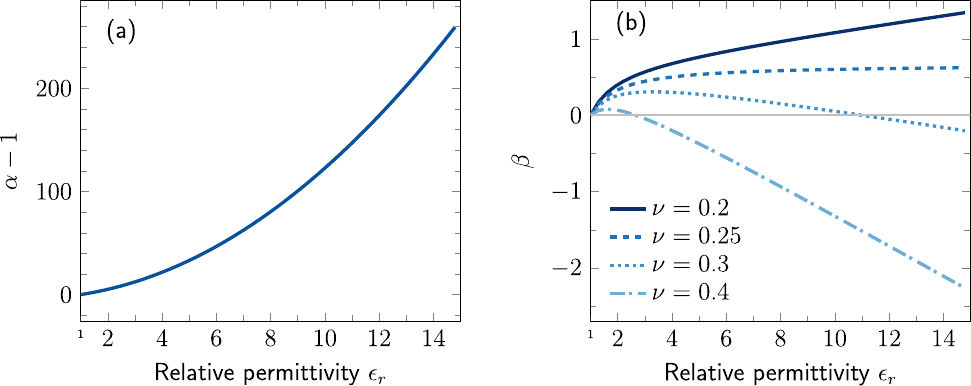}
\caption{Dimensionless parameters as a function of oxide's relative permittivity. (a) The electrostriction parameter $\alpha-1$. (b) The electromechanical coupling parameters $\beta$ at various Poisson's ratios.}\label{fig:alphabeta}
\end{figure}
With all the field quantities, we can get the planar interface velocity
\begin{equation}\label{eq:basic:v0}
\bar{v} = \dod{\bar{h}}{t}=-L\bar{\mathcal{F}},
\end{equation}
where the basic-state driving force is
\begin{equation}\label{eq:Fbar}
\bar{\mathcal{F}} = \Delta f_{\text{chem}} + \bar{\mathcal{E}}_{\text{misfit}} + \bar{\mathcal{E}}_{\text{e}} + \frac{\bar{\mathcal{E}}_{\text{e}}^2}{2(2\mu+\lambda)} + \left(1-\frac{\varepsilon_{\text{s}}}{\varepsilon^*} \frac{\bar{\mathcal{E}}_{\text{misfit}}}{\bar{\mathcal{E}}_{\text{e}}}\right) \alpha_1 \bar{\mathcal{E}}_{\text{e}}^2,
\end{equation}
where $\Delta f_{\text{chem}}$ is the chemical free energy change of oxidation, $\bar{\mathcal{E}}_{\text{misfit}}$ is the strain energy due to the misfit strain given as
\begin{equation}\label{eq:basic:EmisfitEe}
\bar{\mathcal{E}}_{\text{misfit}} = 2\mu\frac{2\mu+3\lambda}{2\mu+\lambda}(\varepsilon^*)^2 = 2\mu \frac{1+\nu}{1-\nu} (\varepsilon^*)^2,
\end{equation}
$\bar{\mathcal{E}}_{\text{e}}=\epsilon \bar{E}_3^2/2$ is the electrostatic energy, $\alpha_1=(\alpha-1)/(2(2\mu+\lambda))$, and $\alpha$ and $\beta$ are two dimensionless parameters defined as
\begin{equation}\label{eq:alphabeta}
\begin{array}{c}
\displaystyle\alpha=\left(1-\frac{a_1}{\epsilon}-\frac{a_2}{\epsilon}\right)\left(1-\frac{3a_1}{\epsilon} - \frac{3a_2}{\epsilon}\right),\\[1em]
\displaystyle\beta = \frac{\lambda a_1/\epsilon-2\mu a_2/\epsilon}{2\mu+\lambda} = \frac{1-2\nu}{1-\nu} \left(\frac{\nu}{1-2\nu}\frac{a_1}{\epsilon} - \frac{a_2}{\epsilon}\right),
\end{array}
\end{equation}
$\varepsilon_{\text{s}}$ is a parameter related to electrostriction
\begin{equation}\label{eq:F:electrostriction:kem}
\varepsilon_{\text{s}} = -\frac{4 \beta}{\alpha-1} \frac{(1-\nu)^2}{(1+\nu)(1-2\nu)},
\end{equation}
where $\nu$ is Poisson's ratio of the oxide film. Equation~\ref{eq:Fbar} shows five contributions to the driving force, including the chemical free energy change $\Delta f_{\text{chem}}$, the strain energy due to misfit strain $\bar{\mathcal{E}}_{\text{misfit}}$, the electrostatic energy $\bar{\mathcal{E}}_{\text{e}}$, the strain energy due to Maxwell stress, and the contribution related to electrostriction. Note that without electrostriction, we have $a_1=a_2=0$, so $\alpha=1$ and $\beta=0$ from Eq.~\ref{eq:alphabeta}, and $\alpha_1=0$ and $\varepsilon_s\alpha_1=0$ from the definition of $\alpha_1$ and Eq.~\ref{eq:F:electrostriction:kem}; therefore, the last term in Eq.~\ref{eq:Fbar} will disappear. For oxidation to occur, we have $\Delta f_{\text{chem}}<0$. The strain energy and electrostatic energy are always non-negative: $\bar{\mathcal{E}}_{\text{misfit}}\geq 0$ and $\bar{\mathcal{E}}_{\text{e}}\geq 0$. Therefore, strain and electrostatic energy always decrease the planar interface thickness (due to a positive $\bar{\mathcal{F}}$). Interestingly, the sign of the electrostriction term can be negative. For the electrostatic-only case (\cite{Du2004}) with interfacial reaction controlled growth, the growth rate is directly related to the electrostatic energy $\sigma = L \left(-k^2 \gamma + 2 \bar{\mathcal{E}}_e k \coth (k\bar{h})\right)$; therefore, we hypothesize that a negative electrostriction term can potentially provide a stabilizing effect. We will demonstrate later that this is indeed the case. The sign of the electrostriction term is negative when
\begin{equation}\label{eq:F:electrostriction}
\frac{\bar{\mathcal{E}}_{\text{e}}}{\bar{\mathcal{E}}_{\text{misfit}}} < \frac{\varepsilon_{\text{s}}}{\varepsilon^*}.
\end{equation}
\begin{figure}[!t]
\centering
\includegraphics[width=.85\textwidth]{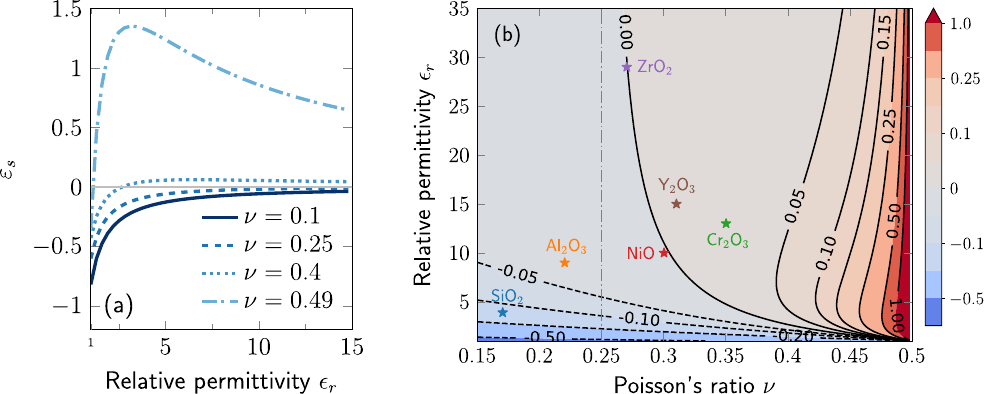}
\caption{Parameter $\varepsilon_s$. (a) Curves at various Poisson's ratios. (b) The $\varepsilon_s$ surface as a function of Poisson's ratio and relative permittivity of the oxide. The markers show typical oxides. The dot-dashed line (Poisson's ratio $\nu=1/4$) is the asymptotic line of the zero contour.}\label{fig:epsilons}
\end{figure}
Figure~\ref{fig:alphabeta} shows the electrostriction parameter $\alpha-1$ and the electromechanical coupling parameter $\beta$ for the model in Eq.~\ref{eq:a1a2}. From Eq.~\ref{eq:F:electrostriction:kem}, we see the sign of $\varepsilon_s$ is determined by parameter $\beta$, since $\alpha-1$ is always non-negative, as shown in Fig.~\ref{fig:alphabeta}a. It is worth noting that parameter $\beta$ characterizes the electromechanical coupling by electrostriction. For example, if we substitute the strain in Eq.~\ref{eq:basic:epsilon33} into the electrical displacement in Eq.~\ref{eq:basic:D3}, the coefficient before the coupling term $\varepsilon^*\bar{E}_3$ is $\beta$. From Fig.~\ref{fig:alphabeta}b, the sign of $\beta$ is always non-negative when $\nu\leq 1/4$, and the resulting $\varepsilon_s$ is always negative, see Fig.~\ref{fig:epsilons}a. For $\nu>1/4$, $\beta$ is initially positive and becomes negative at a sufficiently large $\epsilon_r$ (Fig.~\ref{fig:alphabeta}b), resulting in a positive $\varepsilon_s$ at a large permittivity (Fig.~\ref{fig:epsilons}a). The second case is more interesting as larger permittivity results in a more significant electrostrictive effect. For a positive misfit strain and a large Poisson's ratio, $\varepsilon_s$ can potentially be large enough to fulfill Eq.~\ref{eq:F:electrostriction}, leading to a stabilizing effect by electrostriction. Figure~\ref{fig:epsilons}b shows the parameter $\varepsilon_s$ as a function of Poisson's ratio $\nu$ and relative permittivity $\epsilon_r$, with typical oxides shown by the markers. For typical oxides, $\varepsilon_s$ is negative or small in magnitude; therefore, the electrostriction term acts as a destabilizing factor. However, the region of a large Poisson's ratio is worth exploring for future oxide design.

The case of a zero driving force corresponds to an equilibrium film thickness $\bar{h}_{\text{eq}}$, which is related to an equilibrium electric field through Eq.~\ref{eq:basic:E3}. By setting $\bar{\mathcal{F}}=0$ in Eq.~\ref{eq:Fbar}, we can solve the equilibrium electrical field
\begin{equation}\label{eq:Eeq}
\bar{E}_3^{\text{eq}} = \sqrt{\frac{-(1+2\beta\varepsilon^*)+\sqrt{(1+2\beta\varepsilon^*)^2-4\alpha_2(\Delta f_{\text{chem}} + \bar{\mathcal{E}}_{\text{misfit}})}}{\alpha_2\epsilon}}.
\end{equation}
where $\alpha_2 = \alpha/(2(2\mu+\lambda))$. Note that $\bar{E}_3^{\text{eq}}$ is solely determined by the materials parameters, making it an intrinsic quantity of the system. To have a real solution in Eq.~\ref{eq:Eeq}, we require
\begin{equation}\label{eq:Eeq:constraint}
-\Delta f_{\text{chem}} \geq \bar{\mathcal{E}}_{\text{misfit}},
\end{equation}
which means the magnitude of the chemical free energy drop should be larger than the free energy change due to the misfit. Otherwise, oxidation is not energetically favorable and will not occur. Note that $\bar{h}_{\text{eq}}$ is an unstable equilibrium thickness, as will be shown below.

The derivative of velocity $\bar{v}$ with respect to the film thickness $\bar{h}$ is derived from Eqs.~\ref{eq:basic:E3}, \ref{eq:basic:v0}, and \ref{eq:Fbar} as
\begin{equation}\label{eq:dv0dh}
\dod{\bar{v}}{\bar{h}} 
= \frac{L}{\bar{h}} \left(\epsilon\bar{E}_3^2 + 2\beta\varepsilon^*\epsilon\bar{E}_3^2 + \alpha_2 \epsilon^2 \bar{E}_3^4\right).
\end{equation}
This quantity describes the stability of the planar film. If $\mathrm{d} {\bar{v}}/\mathrm{d} {\bar{h}}<0$, any change in film thickness leads to a velocity that brings the film back to its original thickness, indicating a stable equilibrium. Conversely, when $\mathrm{d} {\bar{v}}/\mathrm{d} {\bar{h}}>0$, the film will be at an unstable equilibrium. Substituting the equilibrium electric field in Eq.~\ref{eq:Eeq} into Eq.~\ref{eq:dv0dh}, we find $\mathrm{d} {\bar{v}}/\mathrm{d} {\bar{h}} > 0$, indicating a film with an equilibrium film thickness is at an unstable equilibrium. This could be understood because thicker film reduces the electric field due to a constant voltage drop across the film (see Eq.~\ref{eq:basic:E3}). Moreover, there is no misfit strain-only term in Eq.~\ref{eq:dv0dh} since the strain in the film due to misfit is independent of the film thickness.

\subsection{Linear stability analysis}\label{sec:method:stability}
We now consider a perturbed interface, as shown in Fig.~\ref{fig:oxidation}a. We follow the standard linear stability analysis procedure given in \cite{Spencer1993}. The profile of the perturbed oxide-fluid interface is described as
\begin{equation}\label{eq:perturbed:h}
h(x,t) = \bar{h}(t) + \hat{h}(t) e^{\iu kx+\sigma t}.
\end{equation}
where $k$ is the wavenumber of the perturbation, $\sigma$ is the growth rate, $\iu$ is the imaginary unit. The normal vector of the perturbed oxide-fluid interface is $\mathbf{n}=\{-\iu k \hat{h} e^{\iu kx+\sigma t},0,1\}^T$, and the total curvature is $\kappa= \hat{\kappa} e^{\iu kx+\sigma t}$ with $\hat{\kappa} = k^2 \hat{h}$.

The solution for the perturbed interface is assumed to be a superposition of the basic-state solution and a perturbed solution as
\begin{equation}\label{eq:perturbed:uphi}
\begin{split}
u_i^{\text{s}}(x,z,t) &= 0 + \hat{u}_i^{\text{s}}(z,t) e^{\iu kx+\sigma t},\\[.5em]
u_i(x,z,t) &= \bar{u}_i(z) + \hat{u}_i(z,t) e^{\iu kx+\sigma t},\\[.5em]
\phi(x,z,t) &= \bar{\phi}(z) + \hat{\phi}(z,t) e^{\iu kx+\sigma t},
\end{split}
\end{equation}
where $\bar{u}_i$ and $\bar{\phi}$ are basic-state solutions given in the previous section. Substituting the perturbation in Eqs.~\ref{eq:perturbed:h} and \ref{eq:perturbed:uphi} into the governing equations in Eqs.~\ref{eq:equilibrium:mechanical} and \ref{eq:equilibrium:electrical}, linearizing in the disturbance quantities, we obtain a system of differential equations for the perturbed displacement $\hat{u}_i$ and perturbed electric potential $\hat{\phi}$ in the oxide film
\begin{equation}\label{eq:perturbed:eq}
\begin{split}
\displaystyle\mu \hat{u}_{1,zz} + \iu k (\mu+\lambda) \hat{u}_{3,z} - 2 k^2 (2\mu+\lambda) \hat{u}_1 = - \iu k (a_1+2 a_2) \frac{\bar{E}_3}{2} \hat{\phi}_{,z},\\[.8em]
\hat{u}_{2,zz}-k^2 \hat{u}_2 = 0,\\[.8em]
\displaystyle(2\mu+\lambda) \hat{u}_{3,zz} + \iu k (\mu+\lambda) \hat{u}_{1,z} - k^2 \mu \hat{u}_3 = k^2 \left(a_1-2\epsilon \right)\frac{\bar{E}_3}{2} \hat{\phi} -\left(a_1+a_2-\epsilon \right) \bar{E}_3 \hat{\phi}_{,zz},\\[.8em]
\displaystyle(a_1+a_2) \bar{E}_3 \hat{u}_{3,zz} + \iu k (a_1+2a_2) \frac{\bar{E}_3}{2} \hat{u}_{1,z} -k^2 a_1 \frac{\bar{E}_3}{2} \hat{u}_3 = \frac{\bar{D}_3}{\bar{E}_3} \left(\hat{\phi}_{,zz} - k^2 \hat{\phi}\right) + k^2 a_1\bar{\varepsilon}_{33} \hat{\phi},
\end{split}
\end{equation}
and in the substrate
\begin{equation}\label{eq:perturbed:eq:substrate}
\begin{split}
\mu^{\text{s}} \hat{u}_{1,zz}^{\text{s}} + \iu k (\mu^{\text{s}}+\lambda^{\text{s}}) \hat{u}_{3,z}^{\text{s}} -2 k^2 (2\mu^{\text{s}}+\lambda^{\text{s}}) \hat{u}_1^{\text{s}} = 0,\\[.5em]
\hat{u}_{2,zz}^{\text{s}}-k^2 \hat{u}_2^{\text{s}} = 0,\\[.5em]
(2\mu^{\text{s}}+\lambda^{\text{s}}) \hat{u}_{3,zz}^{\text{s}} + \iu k (\mu^{\text{s}}+\lambda^{\text{s}}) \hat{u}_{1,z}^{\text{s}} - k^2 \mu^{\text{s}} \hat{u}_3^{\text{s}} = 0.
\end{split}
\end{equation}
Similarly, the boundary conditions in Eqs.~\ref{eq:bc:mechanical:so}-\ref{eq:bc:electrical:phi} are linearized as, at $z=\bar{h}$,
\begin{equation}\label{eq:perturbed:bc}
\begin{split}
\iu k \mu \hat{u}_3 - \iu k \left(\epsilon -\frac{a_1}{2}\right)\bar{E}_3 \hat{\phi } + \mu \hat{u}_{1,z}=\iu k \hat{h}\left(\lambda \bar{\varepsilon }_{33}-(3 \lambda +2 \mu )\varepsilon ^*-\left(\epsilon +a_2\right)\frac{\bar{E}_3^2}{2}\right),\\[.5em]
 \iu k \lambda \hat{u}_1 + (\lambda +2 \mu) \hat{u}_{3,z}-(\epsilon -a_1-a_2)\bar{E}_3\hat{\phi}_{,z}=0,
\\[.5em]
 \hat{\phi} = \hat{h}\bar{E}_3, \\[.5em]
\end{split}
\end{equation}
and at $z=0$
\begin{equation}\label{eq:perturbed:bc2}
\begin{split}
 \iu k \mu \hat{u}_3-\iu k\left(\epsilon -\frac{a_1}{2}\right)\bar{E}_3 \hat{\phi}+\mu \hat{u}_{1,z} = \frac{2 k \mu^{\text{s}} }{\lambda^{\text{s}}+3 \mu^{\text{s}}}\left(\iu \mu^{\text{s}}\hat{u}_3 + (\lambda^{\text{s}}+2 \mu^{\text{s}})\hat{u}_1\right), \\[.5em]
 \iu k \lambda \hat{u}_1+(\lambda +2 \mu) \hat{u}_{3,z}-(\epsilon -a_1-a_2)\bar{E}_3 \hat{\phi}_{,z} = \frac{2 k \mu^{\text{s}}}{\lambda^{\text{s}}+3 \mu^{\text{s}}} \left(-\iu \mu^{\text{s}}\hat{u}_1+ (\lambda^{\text{s}}+2 \mu^{\text{s}})\hat{u}_3\right), \\[.5em]
 \hat{\phi}=0.
\end{split}
\end{equation}
Equations \ref{eq:perturbed:eq}, \ref{eq:perturbed:eq:substrate}, \ref{eq:perturbed:bc} and \ref{eq:perturbed:bc2} can be solved to get the perturbed solutions $\hat{u}_i$ and $\hat{\phi}$ in the oxide film and $\hat{u}_i^{\text{s}}$ in the substrate. The closed-form analytical solutions of $\hat{u}_i$, $\hat{\phi}$ are rather long, so they are provided as a \textit{Mathematica} notebook in the Supplementary Materials. The solution of $\hat{u}_i^{\text{s}}$ is the same as in \cite{Spencer1993} and, therefore, not duplicated here.

Finally, we linearize the interfacial boundary condition in Eq.~\ref{eq:interfacial}, solve the evolution of the magnitude of the perturbation $\hat{h}$, and get the growth rate (not a function of time)
\begin{equation}\label{eq:perturbed:growthrate}
\sigma= \frac{1}{\hat{h}} \dpd{\hat{h}}{t} = -\frac{L}{\hat{h}}\left(\hat{\mathcal{F}}+\gamma\hat{\kappa}\right),
\end{equation}
where the perturbed driving force is
\begin{equation}\label{eq:perturbed:F}
\hat{\mathcal{F}}= \left(\bar{\sigma}_{11}\hat{\varepsilon}_{11}+\bar{\sigma}_{33}\hat{\varepsilon}_{33}\right) + \bar{D}_3\hat{E}_3 + \left(a_2\hat{\varepsilon}_{11}+(a_1+a_2)\hat{\varepsilon}_{33}\right)\bar{E}_3^2,
\end{equation}
where the perturbed strain and electric field are calculated from $\hat{u}_i$ and $\hat{\phi}$
\begin{equation}\label{eq:perturbed:epsilonE}
\hat{\varepsilon}_{11} = \iu k \hat{u}_1,\quad \hat{\varepsilon}_{33}=\dpd{\hat{u}_3}{z},\quad\hat{E}_3=-\dpd{\hat{\phi}}{z}.
\end{equation}
As given in Eq.~\ref{eq:perturbed:h}, a positive growth rate means the magnitude of the perturbation increases with time, indicating an unstable interface, while a negative growth rate indicates a stable interface. From Eq.~\ref{eq:perturbed:growthrate}, interfacial energy always gives a negative growth rate, stabilizing the interface. As shown before, for the electrostatic-only case by \cite{Du2004}, the perturbed driving force is $\hat{\mathcal{F}}\leq 0$. The effect of electrostriction will be studied in detail in the next section. Not surprisingly, it can be shown that the zero-wavenumber growth rate is consistent with the basic-state result $\sigma(k=0) = \partial{\bar{v}}/\partial{\bar{h}}$.

\begin{table}[t!]
\centering
\caption{Materials parameters}\label{tab:matpara}
\begin{tabular}[]{cccl}
\hline
Description & Parameter & Value & Unit\\
\hline
Oxide's shear modulus & $\mu$ & $80$ & \si{\GPa}\\
Oxide's Poisson's ratio & $\nu$ & 0.25 & -\\
Substrate's shear modulus & $\mu^{\text{s}}$ & $80$ & \si{\GPa}\\
Substrate's Poisson's ratio &$\nu^{\text{s}}$ & 0.25 & -\\
Misfit strain & $\varepsilon^*$ & 1\% & -\\
Relative permittivity of oxide &$\epsilon_r$ & 50 & -\\
Chemical driving force & $\Delta f_{\text{chem}}$ & $-1\times10^8$ & \si{\joule\per\meter\cubed}\\
Interfacial energy & $\gamma$ & $0.1$ & \si{\joule\per\meter\squared}\\
\hline
\end{tabular}
\end{table}

\section{Results and discussion}\label{sec:results}
\subsection{Materials parameters}\label{sec:results:materialsparameters}
Table \ref{tab:matpara} lists the materials parameters used in this paper. Lam\'e constant $\lambda$ can be written using Poisson's ratio $\nu$ as $\lambda={2\mu\nu}/{(1-2\nu)}$. The electrostrictive coefficients $a_1$ and $a_2$ are determined from Eq.~\ref{eq:a1a2}.

\subsection{Equilibrium film thickness}\label{sec:results:eq:thickness}
\begin{figure}[!t]
\centering
\includegraphics[width=.8\textwidth]{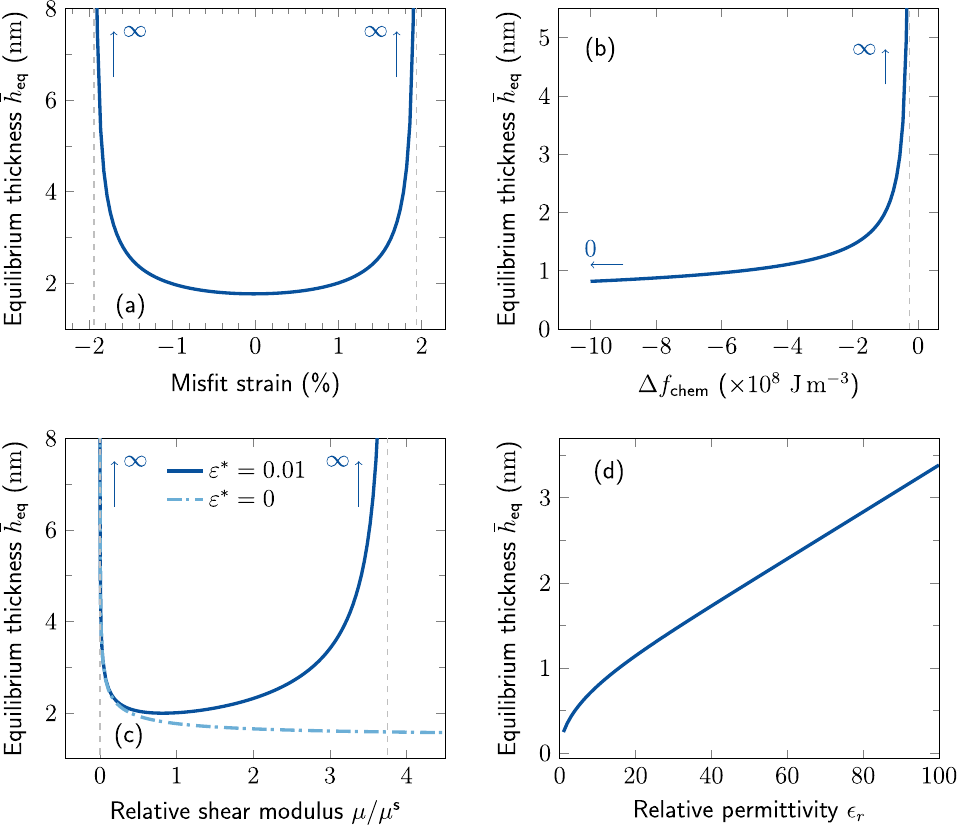}
\caption{Equilibrium film thickness as a function of (a) the misfit strain, (b) the chemical free energy change due to oxidation, (c) the shear modulus of oxide, and (d) the relative permittivity of oxide.}\label{fig:heq}
\end{figure}
Here, we choose $V_{\text{M}}-\Delta\phi^{\text{s-o}}-\Delta\phi^{\text{o-f}}=1~\si{\volt}$. With the materials parameters listed in Table~\ref{tab:matpara}, the equilibrium electrical field calculated from Eq.~\ref{eq:Eeq} is $\bar{E}_3^{\text{eq}}=5\times10^8~\si{\volt\per\meter}$, corresponding to an equilibrium film thickness $\bar{h}_{\text{eq}}=2~\si{\nm}$ using Eq.~\ref{eq:basic:E3}. As shown in Section \ref{sec:method:basic}, this is an unstable equilibrium. However, it should be noted that this equilibrium thickness is only thermodynamically unstable but may be experimentally observable due to kinetic limitations, as discussed in \cite{Du2004}. We anticipate that a steady-state film thickness could be obtained by considering the evolution of both the substrate-oxide and oxide-fluid interface with the oxidation and dissolution reactions and the diffusion and migration of point defects, as in the works of \cite{Macdonald1990,Seyeux2013,Ramanathan2019}. However, this is a complex problem, and the current work focuses on electromechanical coupling; therefore, detailed treatment of diffusion and reactions is not considered here but could be added in the future.

Now, we study the effect of various material parameters on the equilibrium film thickness. We vary one parameter at a time while keeping the rest the same as in Table~\ref{tab:matpara}. Figure~\ref{fig:heq}a shows the equilibrium film thickness as a function of the misfit strain. The equilibrium film thickness is 1.7768~\si{\nm} at zero misfit strain. In this case, the chemical free energy change $\Delta f_{\text{chem}}$ is fully compensated by the energy related to the electric field. As the magnitude of the misfit strain increases, the free energy related to the misfit strain goes up; therefore, the required electric field to give $\bar{\mathcal{F}}=0$ decreases towards zero, and $\bar{h}_{\text{eq}}$ tends to infinity at $\abs{\varepsilon^*}=1.936\%$. For a larger misfit strain, the condition in Eq.~\ref{eq:Eeq:constraint} will not allow a real solution, and thus, the planar interface is never in equilibrium. The effect of the oxidation free energy change is shown in Fig.~\ref{fig:heq}b. For a small $\abs{\Delta f_{\text{chem}}}$, the equilibrium film thickness tends to infinity at $-2.67\times10^{7}~\si{\joule\per\meter\cubed}$. As the magnitude of $\Delta f_{\text{chem}}$ increases, $\bar{h}_{\text{eq}}$ tends to zero. This could be understood by the fact that the electric field required to compensate $\Delta f_{\text{chem}}$ gets bigger as $\Delta f_{\text{chem}}$ approaches $-\infty$; therefore, $\bar{h}_{\text{eq}}$ tends to zero. The effect of oxide's shear modulus is given in Fig.~\ref{fig:heq}c. As the shear modulus tends to zero, the Maxwell stress and electrostrictive terms in Eq.~\ref{eq:Fbar} dominate as they are inversely proportional to $\mu$. The electric field needed to give $\bar{\mathcal{F}}=0$ approaches zero, so $\bar{h}_{\text{eq}}$ tends toward infinity. At the other end, as the shear modulus increases, the strain energy will dominate, so the required electric field tends to zero, resulting in an infinite $\bar{h}_{\text{eq}}$ at $\mu/\mu^{\text{s}}=3.75$. For even larger shear modules, there is no equilibrium film thickness as Eq.~\ref{eq:Eeq:constraint} is not fulfilled. The situation becomes different without the misfit strain $\varepsilon^*=0$, as shown by the dot-dashed line in Fig.~\ref{fig:heq}c. The equilibrium thickness continues to decrease with an increasing $\mu$ since the energy due to misfit $\mathcal{E}_{\text{misfit}}$ remains zero. Figure~\ref{fig:heq}d shows the effect of the oxide's relative permittivity. As the permittivity increases, the electric field required to balance $\Delta f_{\text{chem}}$ decreases, leading to a larger equilibrium film thickness. 

\subsection{Limiting cases}\label{sec:results:limit}
\begin{figure}[!t]
\centering
\includegraphics[width=.9\textwidth]{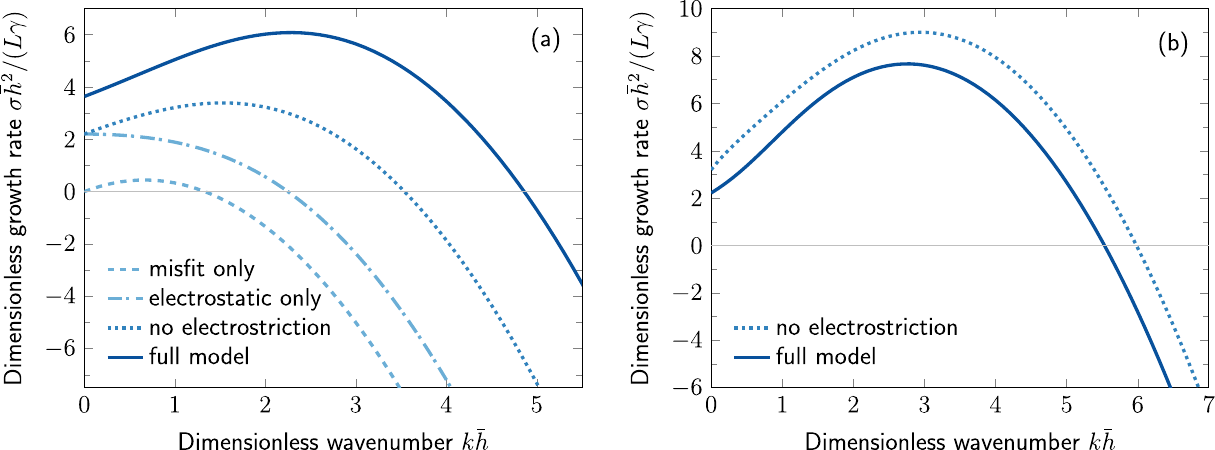}
\caption{Stability dispersion curves. (a) Comparison of stability dispersion curves of various cases. (b) A special case where electrostriction has a stabilizing effect.} \label{fig:result:dispersion:compare}
\end{figure}
It can be shown that our model reduces to simpler models under certain limits. For the first limit, by neglecting the electric field $\bar{E}_3=0$, our model reduces to the instability driven solely by lattice misfit strain given by \cite{Spencer1993}. The second limit is the purely electrostatic field-induced instability by \cite{Du2004}. Our model reduces to this case for zero misfit strain $\varepsilon^*=0$ and a rigid film assumption $\mu\rightarrow\infty$, $\lambda\rightarrow\infty$. This model does not include Maxwell stress, so the rigid film assumption is necessary to eliminate all contributions from the electromechanical coupling. For the last limit, we can neglect the electrostrictive effect by setting $a_1=a_2=0$. Note that there is still an effect through the Maxwell stress, even with a zero misfit strain. Moreover, in the previous work by \cite{Tang2011}, the electric potential is assumed to not change with a perturbation of the interface; therefore, a fully coupled problem is not considered. We find that coupling has a significant effect on the stability result. This model is also a limiting case of our model when assuming zero misfit strain $\varepsilon^*=0$, soft substrate $\mu^{\text{s}}=0$ and $\lambda^{\text{s}}=0$, and zero perturbed electric potential $\hat{\phi}=0$.

With the materials parameters listed in Table~\ref{tab:matpara}, the dispersion curves for different cases for a film of $\bar{h}=\bar{h}_{\text{eq}}=2~\si{\nm}$ thick are shown in Fig.~\ref{fig:result:dispersion:compare}a. The unstable wavenumber range is larger if the electromechanical coupling is included. The maximum growth rate of the full model (solid line, with electrostriction) is more than one order of magnitude larger (13.7 times) than the misfit-only case (dashed line) and 2.8 times larger than the electrostatic-only case (dot-dashed line). This indicates that neglecting the electromechanical coupling may lead to an overly optimistic estimate of the stability of the planar film. For the specific set of materials parameters used here, the electrostrictive term in Eq.~\ref{eq:Fbar} is positive, consistent with the result in Fig.~\ref{fig:result:dispersion:compare}a: a destabilizing effect by electrostriction.

In Section~\ref{sec:method:basic}, we showed that electrostriction could have a stabilizing effect like interfacial energy. To demonstrate this, we change the Poisson's ratio to $\nu=0.49$ to fulfill Eq.~\ref{eq:F:electrostriction}. The resulting dispersion curve of the full model (solid line) is compared with the one without electrostriction (dotted line) in Fig.\ref{fig:result:dispersion:compare}b. The critical wavenumber (non-zero wavenumber where the growth rate curve crosses zeros) is smaller if electrostriction is considered, confirming a stabilizing effect by electrostriction. Although typical oxides do not have such a large Poisson's ratio, this provides a possible way to utilize electrostriction to stabilize passive oxide films.

\subsection{Effect of parameters on stability}\label{sec:results:study}
\begin{figure}[!t]
\centering
\includegraphics[width=.9\textwidth]{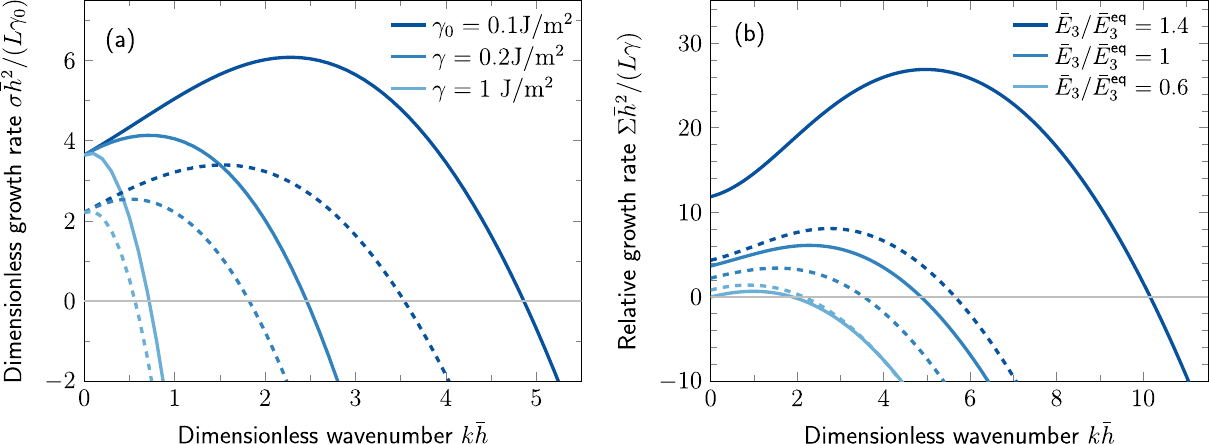}
\caption{Stability dispersion curves. (a) The effect of interfacial energy $\gamma$ with solid lines shows the results with electrostriction and dashed lines without electrostriction. The growth rate is nondimensionalized by $\gamma=0.1~\si{\joule\per\meter\squared}$. (b) The effect of the electric field $\bar{E}_3$ on the relative growth rate $\Sigma=\sigma-\bar{v}/\bar{h}$. The equilibrium electric field is determined from the parameters in Table~\ref{tab:matpara}. Dashed curves are results without electrostriction.}\label{fig:result:effect:GammaAndE}
\end{figure}
To investigate the system's stability, we vary the model parameters and study the effect on the growth rate. The materials parameters listed in Table~\ref{tab:matpara} are used except for the varied one. 

The effect of interfacial energy on the stability of a $\bar{h}=\bar{h}_{\text{eq}}=2~\si{\nm}$ thick oxide film is shown in Fig.~\ref{fig:result:effect:GammaAndE}a. Since the interfacial energy stabilizes the film, reducing the interfacial energy leads to a larger perturbation growth rate and a wider range of unstable wavenumbers. Moreover, the change in the unstable range is larger than the case without electrostriction (dashed lines in Fig.~\ref{fig:result:effect:GammaAndE}a). In practice, the interfacial energy can be tailored by additives in the solution. Decreasing interfacial energies by adding chloride (salt) will lead to a more unstable interface and hence less protection.

Next, we examine the effect of the electric field. We fix the film thickness to $\bar{h}=2~\si{\nm}$ while varying the electric field by $V_{\text{M}}$. The resulting electric field $\bar{E}_3$ deviates from the equilibrium electric field $\bar{E}_3^{\text{eq}}=0.5~\si{\volt\per\nm}$ determined from Eq.~\ref{eq:Eeq}. Therefore, the planar interface is not stationary. To account for a non-zero $\bar{v}$, we use the relative growth rate defined by \cite{Spencer1993}
\begin{equation}\label{eq:relative:sigma}
\Sigma = \left(\frac{\hat{h}}{\bar{h}}\right)^{-1} \dpd{}{t} \left(\frac{\hat{h}}{\bar{h}}\right) = \sigma - \frac{\bar{v}}{\bar{h}}.
\end{equation}
The stability dispersion curves for various electric field values are shown in Fig.~\ref{fig:result:effect:GammaAndE}b. For an increase in the electric field of 40\% from $\bar{E}_3^{\text{eq}}$, the range of unstable wavelengths changes by about a factor of two, but the maximum growth rate increases by a factor of 4.4. Similar behavior is observed for a 40\% decrease from $\bar{E}_3^{\text{eq}}$, where the range of unstable wavelengths changes by a factor of 2.6 and the maximum growth rate decreases by a factor of 9.4. This is far more sensitive than purely mechanical effects. By comparing with the results without electrostriction (dashed curves in Fig.~\ref{fig:result:effect:GammaAndE}b), for the materials parameters used here (Table~\ref{tab:matpara}), electrostriction leads to positive feedback between strain and polarization, causing a more unstable film that is sensitive to the electric field. For the analysis below, the linear stability analysis is performed on the equilibrium state of the planar interface, \textit{i.e.} $\bar{E}_3=\bar{E}_3^{\text{eq}}$, $\bar{v}=0$, and $\Sigma=\sigma$.

\begin{figure}[!t]
\centering
\includegraphics[width=.9\textwidth]{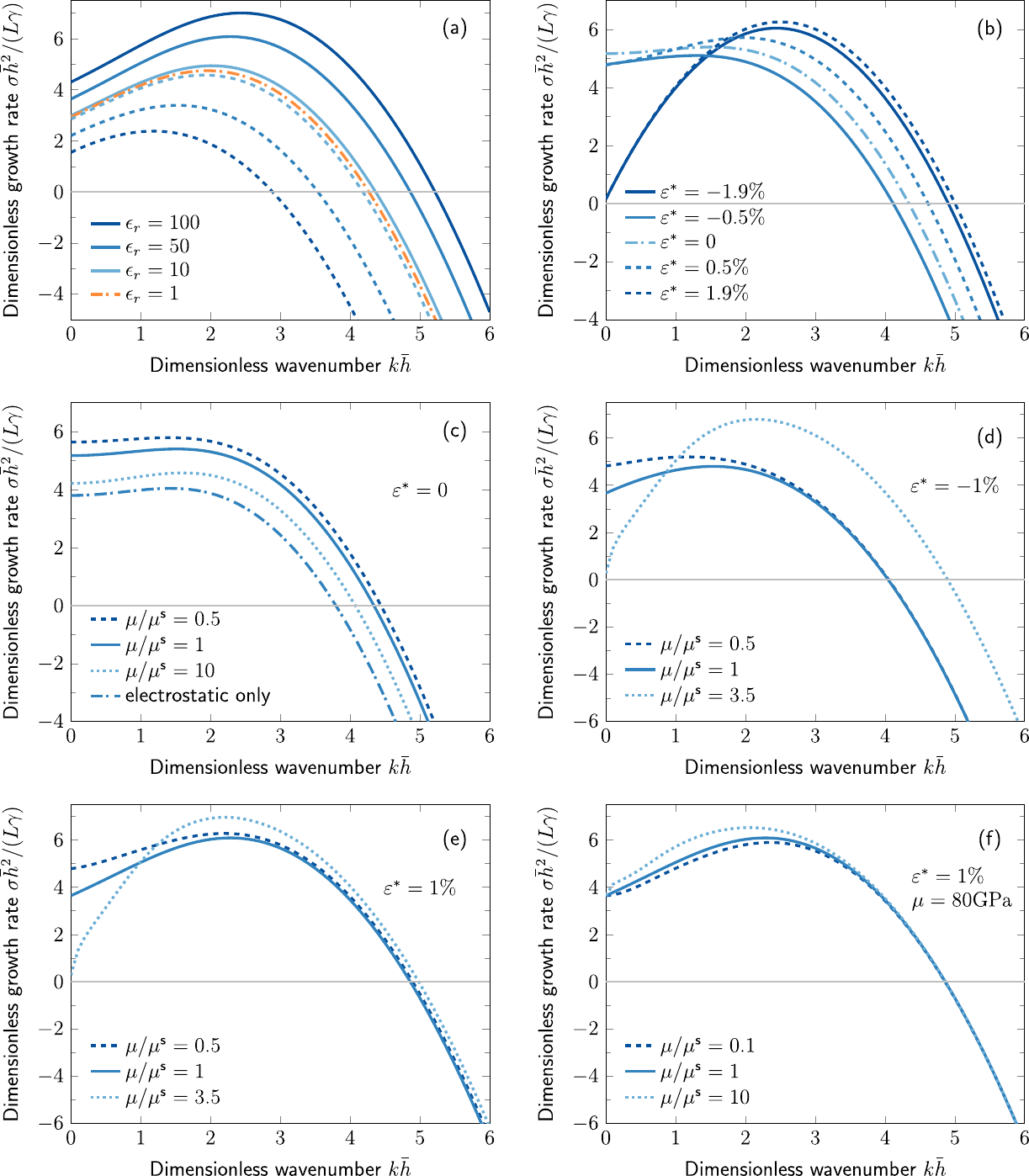}
\caption{Stability dispersion curves of the equilibrium state. (a) The effect of oxide's relative permittivity. Dashed lines are results without electrostriction with varying $\epsilon_r$ as denoted by the color. The dot-dashed line shows the result of $\epsilon_r=1$, equivalent to no-electrostriction. (b) The effect of misfit strain. Solid and dashed lines are results with and without electrostriction, respectively. The dot-dashed line shows the result of zero misfit. (c-e) The effect of oxide's shear modulus ($\mu^s=80~\si{\GPa}$ fixed) for a zero-misfit (c), a negative misfit (d), and a positive misfit (e). (f) The effect of substrate's shear modulus ($\mu=80~\si{\GPa}$ fixed).}\label{fig:result:effects}
\end{figure}

Now we investigate the influence of oxide's relative permittivity by varying within the range of permittivities for typical oxides (\textit{e.g.}, $\epsilon_r=9$ for Al$_2$O$_3$, $\epsilon_r=95$ for TiO$_2$ (\cite{McPherson2003})). The results are shown in Fig.~\ref{fig:result:effects}a by the solid lines for cases with electrostriction and dashed lines for cases without. For a fixed electric field, the oxide film is less stable with a large permittivity (due to a large electrostatic energy $\bar{\mathcal{E}}_e$). Here, the electric field is fixed to be the equilibrium one, which varies with permittivity, as shown in Fig.~\ref{fig:heq}d (note $\bar{E}_3^{\text{eq}}\propto \bar{h}_{\text{eq}}^{-1}$). In contrast to the case with electrostriction, for the no-electrostriction case (dashed lines), larger permittivity gives a more stable film. This is due to a decrease in $\bar{E}_3^{\text{eq}}$ with increasing permittivity. However, with electrostriction, even if the equilibrium electric field is smaller for higher permittivity, the net effect of permittivity is destabilization. In the limit of $\epsilon_r=1$, the electrostriction disappears ($a_1=a_2=0$), so the electrostriction curve (solid line) and the no-electrostriction curve (dashed line) collapse, which is shown by the dot-dashed line in Fig.~\ref{fig:result:effects}a.

We now move to study the effect of the misfit strain $\varepsilon^*$. The sign of the misfit strain can be positive or negative, depending on the structures of the oxide and substrate. The dispersion curves for various misfit strains are shown in Fig.~\ref{fig:result:effects}b. The growth rate at zero wavenumber (Eq.~\ref{eq:dv0dh}) decreases with increasing absolute value of the misfit strain, which is mainly due to a change in the equilibrium electric field, as shown in Fig.~\ref{fig:heq}a (note $\bar{E}_3^{\text{eq}}\propto \bar{h}_{\text{eq}}^{-1}$). In the misfit-only case, a larger magnitude of misfit strain leads to unstable film (due to larger strain energy $\bar{\mathcal{E}}_{\text{misfit}}$), and the sign of the misfit strain does not affect the stability (\cite{Spencer1993}). However, electrostriction leads to a non-symmetric dependence on the misfit strain, as shown in Fig.~\ref{fig:result:effects}b by comparing the solid and dashed lines of the same color. Although $\bar{\mathcal{E}}_{\text{misfit}}$ is symmetric with respect to $\varepsilon^*$, the electrostrictive part depends on the sign of $\varepsilon^*$. Overall, the growth rate does not change significantly, indicating a small effect of misfit strain on a film at the equilibrium state.

We now look at the effect of the relative stiffness, \textit{i.e.}, the ratio between the oxide's shear modulus $\mu$ and the substrate's shear modulus $\mu^{\text{s}}$. The dispersion curves for zero, negative, and positive misfit strains are shown in Figs.~\ref{fig:result:effects}c, \ref{fig:result:effects}d, and \ref{fig:result:effects}e, respectively. For zero misfit strain, $\bar{\mathcal{E}}_{\text{misfit}}=0$, but Maxwell and electrostrictive stresses still contribute to the strain energy. As shown in Fig.~\ref{fig:result:effects}c, a larger oxide's shear modulus leads to a more stable film without the misfit strain. This is because the contribution from the electromechanical coupling reduces with an increasing $\mu$, as Maxwell and electrostrictive parts are proportional to $\mu^{-1}$, see Eq.~\ref{eq:Fbar}. The dispersion curve converges to the electrostatic-only case (dot-dashed line) as $\mu$ approaches infinity. For a non-zero misfit strain, there is an upper bound on the maximum shear modulus $\mu$ to permit an equilibrium electric field, as shown in Fig.~\ref{fig:heq}c. The equilibrium electric fields for $\mu/\mu^{\text{s}}=0.1$ and $\mu/\mu^{\text{s}}=1$ are similar but reduce by a factor of 2.9 for $\mu/\mu^{\text{s}}=3.5$. However, the net effect of electrostriction is destabilizing. The film is less stable with a larger $\mu$ for a non-zero misfit strain, although the effect is small, especially for a positive misfit. Recall from Fig.~\ref{fig:epsilons}a that $\varepsilon_s<0$ for the materials parameters used here ($\nu=0.25$). The electrostriction part in Eq.~\ref{eq:Fbar} has a smaller magnitude for a negative misfit strain $\varepsilon^*<0$ than a positive misfit one $\varepsilon^*>0$, leading to a more stable film with a negative misfit. This can be seen by comparing the dispersion curves in Fig.~\ref{fig:heq}d and Fig.~\ref{fig:heq}e.

Next, we keep $\mu$ constant and vary $\mu^{\text{s}}$. In this case, the equilibrium electric field is unchanged. As shown by the dispersion curves in Fig.~\ref{fig:result:effects}f, the shear modulus of the substrate has little effect on the film's stability (unlike in the misfit-only case). This is because the film's strain energy change due to the substrate's shear modulus is small compared to the electromechanical coupling.

\section{Conclusions}\label{sec:conclusion}
We have presented a continuum model accounting for electromechanical coupling by electrostriction to understand the critical role of electrostriction in the growth and breakdown of passivation films. The model includes a full coupling between mechanical deformation and electric polarization. More importantly, the model also considers lattice misfit, Maxwell stress, and electrostriction.  We apply the model to study the stability of thin oxide films and perform morphological linear instability analysis. We find that with a large electric field, common in thin oxide films, electrostriction is critical in determining the morphological stability of the film. Neglecting the electrostrictive effects can lead to a poor estimate of when the passivation film becomes morphologically unstable. Through the analysis, we identified an equilibrium electric field, an intrinsic property of the system. The equilibrium electric field corresponds to an equilibrium film thickness at which a planar oxide interface is stationary. This equilibrium film thickness is thermodynamically unstable but can be stable if the oxidation process is limited by kinetics. The film's stability is found to be very sensitive to the electric field. A slight deviation from the equilibrium electric field leads to a significant change in the film's stability. So, one efficient way to alter the film's stability is by tuning the equilibrium electric field through materials parameters design. Unlike pure electrostatic instability, which acts only as destabilizing, our analysis identifies a parameter region (large Poisson's ratio, positive misfit, moderate to large permittivity) where electrostriction can provide a stabilizing effect like the interfacial energy. Although not applicable to typical oxides, it potentially provides a new mechanism for designing new oxides with improved corrosion resistance. Finally, through parameter study, we find that the morphological stability of the film can be improved by increasing the interfacial energy and decreasing the electric field. The substrate's shear modulus is found to have a negligible effect on the stability. The effect of oxide's permittivity, misfit strain, and oxide's stiffness on the film's stability at the equilibrium electric field is complicated by electrostriction (\textit{e.g.}, symmetry breaking of the misfit strain) but is generally found to have a negligible effect at the equilibrium state.

The oxidation model presented here serves as a first step toward understanding the influence of electromechanical coupling on the passivation of thin oxide films. The study assumes the growth of oxide is reaction-controlled. For diffusion-controlled growth, detailed transport of defects within the oxide film should be considered. See, for example, the point-defect model for oxide film growth (\cite{Macdonald1992}) and others as discussed in the work of \cite{Ramanathan2020}. The stability of both the oxide-fluid and the metal-oxide interfaces could also be considered (\cite{Hebert2012,Sato1971,Singh2006}), depending on the primary diffusing species (\cite{Ramanathan2019}).

\section*{Acknowledgments}
The authors gratefully acknowledge financial support from the Ofﬁce of Naval Research (ONR) [grant number N00014-16-1-2280]. Helpful discussion with Rohit Ramanathan, Laurence Marks and Brain Spencer is gratefully acknowledged.

\appendix
\section{Derivation of the interfacial boundary condition}\label{sec:appendix:interfacial}
The interfacial boundary condition can be found in \cite{Tang2011,Suo2010}, but a derivation is missing. Therefore, here, we provide a derivation of the interfacial condition of the oxide-fluid interface following the variational approach by \cite{Voorhees2004}. Considering a system composed of two phases (oxide $V^{\text{o}\prime}$ and fluid $V^{\text{f}}$) separated by an interface $\Sigma'$, the total electric Gibbs free energy of the system is
\begin{equation}\label{eq:G}
\mathcal{G} = \intj{V^{\text{o}'}}{g_{v'}^{\text{o}}({F_{ij}},{E}_i^{\text{o}\prime})}{V} + \intj{V^{\text{f}}}{g_{v}^{\text{f}}({E}_i^{\text{f}})}{V} + \intj{\Sigma'}{g_{\text{a}'}^\Sigma(\kappa_1',\kappa_2')}{A} + \intj{\Sigma'}{w_e \phi^\Sigma}{A},
\end{equation}
where $g_{v'}^{\text{o}}$ and $g_{v}^{\text{f}}$ are the free energy density of the oxide and fluid phases, respectively, $g_{\text{a}'}^\Sigma$ is the free energy density per reference area, $w_e$ is the surface charge density, ${F}_{ij}$ is the deformation gradient, ${E}_i^{\text{o}\prime}$ is the nominal electric field (per area in the reference state) in the oxide, ${E}_i^{\text{f}}$ is the true electric field (per current area) in the fluid, $\kappa_a'$ and $\kappa_2'$ are principal curvatures in the reference state, $\phi^\Sigma=(\phi^{\text{o}}+\phi^{\text{f}})/2$. Note that we use $(\cdot)'$ to denote quantities in the reference configuration. The electric Gibbs free energy is defined as the Legendre transform of the Gibbs free energy $f$:
\begin{equation}\label{eq:g2}
g({F}_{ij},{E}'_i) = f({F}_{ij},{D}'_i) - {D}'_k{E}'_k,
\end{equation}
where ${D}'_i$ is the nominal electric displacement field.

Variations in the energy densities under isothermal and constant pressure conditions are
\begin{equation}\label{eq:deltag}
\begin{array}{l}
\delta g_{v'}^{\text{o}} = {P}_{ji}^{\text{o}}\delta{F}_{ij} - {D}_i^{\text{o}\prime}\delta{E}_i^{\text{o}\prime},\\[.5em]
\delta g_v^{\text{f}} = - {D}_i^{\text{f}}\delta{E}_i^{\text{f}},\\[.5em]
\delta g_{a'}^\Sigma = K_1 \delta \kappa_1' + K_2 \delta \kappa_2'.
\end{array}
\end{equation}
where ${P}_{ij}^{\text{o}}$ is the first Piola-Kirchhoff stress, including Cauchy and Maxwell stresses. The reader can refer to the work by \cite{Suo2008,Suo2010} for a general discussion on Maxwell stress.

Here, we neglect the surface stress and assume the interfacial double layer does not vary with the variations. We can choose the dividing plane to eliminate the variation with $\delta \kappa_1'$ and $\delta \kappa_2'$ (\cite{Voorhees2004}).

In calculating the variation of $\mathcal{G}$ in Eq.~\ref{eq:G}, we must account for the change of the integral domain due to phase transformation at the interface $\Sigma'$. The extent of phase transformation is tracked relative to the reference state of the oxide crystal and measured by an accretion vector $\delta y^{\text{o}\prime} {n}_i^{\text{o}\prime}$, where ${n}_i^{\text{o}\prime}$ is the unit normal vector of $\Sigma'$ pointing from the oxide to the fluid. An infinitesimally small variation of a field $f'$ defined in the oxide that, due to phase transformation, can be calculated as
\begin{equation}\label{eq:deltag:1}
\delta \intj{V^{\text{o}\prime}}{f'}{V} = \intj{V^{\text{o}\prime}}{\delta f'}{V} + \intj{\Sigma'}{f' \delta y^{\text{o}\prime}}{A}
\end{equation}
Similarly, the variation of a field $f$ defined in the fluid as a result of phase transformation is calculated as
\begin{equation}\label{eq:deltag:2}
\delta \intj{V^{\text{f}}}{f}{V} = \intj{V^{\text{f}}}{\delta f}{V} + \intj{\Sigma}{f \delta y^{\text{f}}}{A}
\end{equation}
where $\delta y^{\text{f}}$ is the accretion to the fluid in the current state. Since accretion to the crystal varies the surface area measured in the reference state, the variation of the surface integral is calculated as
\begin{equation}\label{eq:deltag:3}
\delta \intj{\Sigma'}{f}{A} = \intj{\Sigma'}{\delta f}{A} + \intj{\Sigma'}{f \kappa' \delta y^{\text{o}\prime}}{A}
\end{equation}
where $\kappa'=\kappa_1'+\kappa_2'$ is the total curvature of $\Sigma'$.

With the help of these results, the variation of the total energy in Eq.~\ref{eq:G} is
\begin{equation}\label{eq:delta:G}
\begin{array}{rl}
\delta \mathcal{G} &\displaystyle= \intj{V^{\text{o}\prime}}{\delta g_{v'}^{\text{o}}}{V} + \intj{V^{\text{f}}}{\delta g_v^{\text{f}}}{V} + \intj{\Sigma'}{g_{v'}^{\text{o}} \delta y^{\text{o}\prime}}{A} + \intj{\Sigma}{g_v^{\text{f}} \delta y^{\text{f}}}{A} \\[.7em]
&\displaystyle+ \intj{\Sigma'}{\delta g_{a'}^\Sigma + g_{a'}^\Sigma \kappa' \delta y^{\text{o}\prime}}{A} + \intj{\Sigma'}{w_e \delta \phi^\Sigma}{A}
\end{array}
\end{equation}
Substituting Eq.~\ref{eq:deltag} into Eq.~\ref{eq:delta:G}, and using $\delta {F}_{ij} = \partial{(\delta{u}_i)}/\partial{X_j}$, and $\delta{E}'_i = -\partial{(\delta\phi)}/\partial{X_i}$ ($X_i$ is the coordinate in the reference state), we have
\begin{equation}\label{eq:delta:G:2}
\begin{array}{rl}
\delta \mathcal{G} &\displaystyle=\intj{V^{\text{o}\prime}}{\left( - (\partial_i{{P}_{ij}^{\text{o}}})\delta{u}_j - (\partial_i{{D}_i^{\text{o}\prime}})\delta\phi^{\text{o}} \right)}{V} + \intj{V^{\text{f}}}{ - (\partial_i{{D}_i^{\text{f}}})\delta\phi^{\text{f}}}{V}\\[.7em]
&\displaystyle+\intj{\Sigma'}{\left( g_{v'}^{\text{o}} \delta y^{\text{o}\prime} + {n}_i^{\text{o}\prime}{P}_{ij}^{\text{o}}\delta {u}_j + {n}_i^{\text{o}\prime}{D}_i^{\text{o}\prime}\delta\phi^{\text{o}} \right)}{A}\\[.7em]
&\displaystyle+\intj{\Sigma}{\left( g_{v}^{\text{f}} \delta y^{\text{f}} + {n}_i^{\text{f}}{D}_i^{\text{f}}\delta\phi^{\text{f}} \right)}{A} +\intj{\Sigma'}{\gamma' \kappa' \delta y^{\text{o}\prime}}{A} + \intj{\Sigma'}{w_e \delta \phi^\Sigma}{A},
\end{array}
\end{equation}
where $\gamma'=g_{a'}^\Sigma$. We need to account for the constraints at the interface. First, the elastic displacement of the two phases at the interface is continuous, so the two phases are constantly in contact. From this, the variation in the displacement of the interface is related to the condition that the oxide and fluid must be in contact in the varied state:
\begin{equation}\label{eq:deltay}
-\delta y^{\text{f}} = {n}_i^{\text{o}}\left(\delta {u}_i + {F}_{ij}{n}_j^{\text{o}\prime} \delta y^{\text{o}\prime}\right).
\end{equation}
Second, we assume the electric potential jump at the interface is constant, so the variation of the electric potential at the interface is
\begin{equation}\label{eq:delta:phi:surface}
\delta\phi^\Sigma = \delta \phi^{\text{o}} - {E}_i^{\text{o}\prime}{n}_i^{\text{o}\prime}\delta y^{\text{o}\prime} = \delta \phi^{\text{f}} - {E}_i^{\text{f}}\left({n}_i^{\text{f}}\delta y^{\text{f}}-\delta {u}_i\right).
\end{equation}
Using these two conditions, we have
\begin{equation}\label{eq:delta:G:3}
\begin{array}{rl}
\delta \mathcal{G} &\displaystyle=\intj{V^{\text{o}\prime}}{ - (\partial_i{{P}_{ij}^{\text{o}}})\delta{u}_j - (\partial_i{{D}_i^{\text{o}\prime}})\delta\phi^{\text{o}}}{V} + \intj{V^{\text{f}}}{ - (\partial_i{{D}_i^{\text{f}}})\delta\phi^{\text{f}}}{V}\\[.7em]
&\displaystyle+\intj{\Sigma'}{\left(g_{v'}^{\text{o}} -J g_v^{\text{f}} + {n}_i^{\text{o}\prime}(J({F}^{-1})_{ik}{D}_k^{\text{f}})({E}_j^{\text{o}\prime} - {E}_l^{\text{f}}{F}_{lj}){n}_j^{\text{o}\prime}- w_e {E}_i^{\text{o}\prime}{n}_i^{\text{o}\prime} + \gamma' \kappa'\right) \delta y^{\text{o}\prime}}{A}\\[.7em] 
&\displaystyle+\intj{\Sigma'}{{n}_i^{\text{o}\prime}({P}_{ij}^{\text{o}}- g_v^{\text{f}} J ({F}^{-1})_{ij})\delta {u}_j}{A}\\[.7em] 
&\displaystyle+\intj{\Sigma'}{\left({n}_i^{\text{o}\prime}({D}_i^{\text{o}\prime}-J({F}^{-1})_{ik}{D}_k^{\text{f}})+w_e\right)\delta \phi^{\text{o}}}{A},
\end{array}
\end{equation}
where $J=\det(F_{ij})$. The equilibrium condition $\delta\mathcal{G}=0$ is, thus,
\begin{align}\label{eq:delta:G:equilibrium}
    \dpd{{P}_{ij}^{\text{o}}}{X_i} = 0, \quad \dpd{{D}_i^{\text{o}\prime}}{X_i}=0, \quad \dpd{{D}_i^{\text{f}}}{x_i}=0,\nonumber\\
    {n}_i^{\text{o}\prime}({P}_{ij}^{\text{o}} - g_v^{\text{f}} J ({F}^{-1})_{ij})= 0,\nonumber\\
    {n}_i^{\text{o}\prime}({D}_i^{\text{o}\prime} - {D}_j^{\text{f}} J({F}^{-1})_{ij}) = -w_e.
\end{align}
In this work, we assume interfacial reaction is rate-limiting, so the interface should evolve towards minimizing the total energy $\mathcal{G}$. For linear kinetics, we have the interfacial velocity
\begin{equation}\label{eq:appd:vn}
v_n = \dod{y^{\text{o}}}{t}=-L\dfd{\mathcal{G}}{y^{\text{o}}},
\end{equation}
where the functional derivative is obtained by substituting the expression for $w_e$ into Eq.~\ref{eq:delta:G:3} and using the definitions $g_{v'}^{\text{f}}=J g_v^{\text{f}}$, ${{D}}_i^{\text{f}\prime}=J({F}^{-1})_{ik}{D}_k^{\text{f}}$ and ${{E}}_i^{\text{f}\prime}={E}_k^{\text{f}}{F}_{ki}$:
\begin{equation}\label{eq:appd:dG}
\dfd{\mathcal{G}}{y^{\text{o}\prime}}=g_{v'}^{\text{o}} -g_{v'}^{\text{f}} + {n}_i^{\text{o}\prime} \left({D}_i^{\text{o}\prime}{E}_j^{\text{o}\prime} - {{D}}_i^{\text{f}\prime}{{E}}_j^{\text{f}\prime}\right){n}_j^{\text{o}\prime} + \gamma' \kappa' = 0.
\end{equation}
This is the expression for the interfacial condition. In the limit of small strain, there is no need to distinguish the prime and non-prime quantities. Moreover, we assume a zero electric field in the fluid. The interfacial condition can be simplified as
\begin{equation}\label{eq:appd:vn:2}
v_n =-L\left(g_{v}^{\text{o}}-g_{v}^{\text{f}} + {n}_i^{\text{o}} {D}_i^{\text{o}}{E}_j^{\text{o}}{n}_j^{\text{o}}+ \gamma \kappa\right).
\end{equation}

% ======================================================================

\end{document}